\documentclass[journal=apchd5,manuscript=article]{achemso}

\usepackage{graphicx}

\usepackage[T1]{fontenc} 
\usepackage{chemformula} 

\author{Aaswath Raman}
\affiliation[UCLA]{Department of Materials Science and Engineering, University of California, Los Angeles, Los Angeles, California 90049, USA}
\email{aaswath@ucla.edu}
\author{Zongfu Yu} 
\affiliation[UWisc]{Department of Electrical and Computer Engineering, University of Wisconsin, Madison, Wisconsin 53706, USA}
\author{Shanhui Fan}
\affiliation[Stanford]{Department of Electrical Engineering, Ginzton Laboratory, Stanford University, Stanford, California 94305, USA}

\title[Limits]
  {Limits on Broadband Absorption Enhancement in the Presence of Multiple Lossy Materials}

\begin{document}

\begin{abstract}
Enhancing the absorption and emission of electromagnetic waves over a broad range of wavelengths is a topic of fundamental and applied interest in photonics and energy research. In the context of light trapping in solar cells, for example, significant interest in the past decade has focused on overcoming limits in the ray-optics regime with nanophotonic structures. However, many such structures, in particular plasmonic structures, or $PT$-symmetric systems can posses multiple materials with varying values of intrinsic loss. Here, we rigorously determine the effect of parasitic loss on the achievable absorption enhancement in arbitrary electromagnetic structures. We show that the fundamental limit of broadband absorption enhancement, even in the presence of large parasitic loss in an alternate material, can exceed conventional ray-optics limits on light trapping and absorption enhancement. We numerically verify this behavior by determining the absorption enhancement factor of a canonical system, a metal-insulator-metal waveguide, whose core is a low-index organic semiconductor, in the presence of varying intrinsic loss values in the metal.
\end{abstract}
\maketitle
\noindent

\section{Introduction}
Enhancing broadband light absorption in thin, sub-wavelength layers of materials is a topic of fundamental interest in contemporary materials, photonics and photovoltaics research. Following early work in light trapping for solar cells\cite{{Yab82},{yablonovitch1982},{StuartHall}}, a recent wave of work has sought to apply nanophotonic techniques to enhancing light absorption in thin active layers for next-generation solar cells\cite{{zhu2009},{Mallick10},{garnett2010},{ferry2010},{polman2012},{battaglia2012},{grote2013},{narasimhan2013},{brongersma2014}}. Related to the wide array of systems and devices considered, the theoretical framework required to understand the fundamental limit of absorption enhancement was concurrently extended beyond the ray optics approximation, to account for wavelength- and subwavelength-scale effects present in nanoscale active layers \cite{{YuP},{YuOE},{Callahan2012},{Yu2012},{schiff2011},{munday2012},{mokkapati2012},{buddhiraju2017}}. This nanophotonic light trapping theory\cite{{YuP},{Yu2012}} established mechanisms by which one can exceed the conventional limit on the absorption enhancement factor $F$ of $4n^2$, where $n$ is the bulk refractive index of the bulk absorber for which conventional limits do apply. 

One particular category of nanophotonic absorption enhancement schemes that has been extensively investigated is the use of metallic nanostructures, which support plasmonic modes at deep subwavelength scales, to enhance light absorption in a thin absorber \cite{{catchpole2008},{kim2008plasmon},{beck2009},{atwater2010},{ferry2010design},{Green2012},{spinelli2012}}. It has been shown that in the case of ultra-thin absorbers, that this approach offers the possibility of exceeding the conventional broadband $4n^2$ limit, where $n$ is the refractive index of the absorbing material \cite{munday2012}. However, a key challenge with the use of metallic nanostructures for light trapping has been parasitic absorption of light in the metal. Many numerical and analytical analyses have been undertaken on specific plasmonic light trapping structures where the loss in the metal has been taken into account\cite{{pala2009},{pillai2010},{ferry2011modeling},{tan2012},{pala2013},{holman2014},{morawiec2016},{disney2017}}. Theoretical works on limits to light trapping\cite{{schiff2011},{sheng2012},{Green2012}} on the other hand have primarily focused on assuming that the active semiconductor layer to be the only absorbing layer present, with analyses of parasitic loss considered for specific scenarios and approximations. A rigorous, universal analysis of the impact of parasitic loss on the fundamental limits of absorption enhancement remains to be developed. 

Beyond metallic nanostructures, a better understanding of light absorption or emission in subwavelength volumes in the presence of multiple intrinsic loss mechanisms is a topic of broad relevance. All solar cells in practice have lossy regions, such as electrodes or heavily doped regions in semiconductor cells, which are essential for the operation of the cell, but where absorption of light does not contribute to photocurrent. By Kirchhoff's law, understanding how broadband absorption enhancement behaves in the presence of multiple lossy materials also reveals how thermal emission might be enhanced in such complex photonic structures. From thermal radiation applications such as radiative cooling\cite{Raman2014} and thermophotovoltaics, to the use of high-index nanostructures in photodetectors and solar cells, there is thus a general need to understand how the ability to enhance broadband absorption and emission in subwavelength volumes is affected by the presence of non-zero absorption in non-active materials in, or near, the active volume.

In this paper we develop and evaluate a formalism that describes how the introduction of multiple lossy materials influences the broadband absorption enhancement limit in nanophotonic structures and metamaterials. This formalism is in general applicable for any system with multiple lossy materials, and explicitly accounts for the impact of parasitic absorption in a non-active material, and for any active layer volume. We focus in particular on how parasitic absorption from metal influences the capability of plasmonic light trapping structure to exceed the conventional limit of broadband absorption enhancement. We rigorously show that in the weak active-layer absorption regime, parasitic absorption reduces the achievable absorption enhancement factor $F_p$ in a desired material or region of a nanophotonic structure. However, with appropriate design choices, this lowered enhancement factor $F_p$ can still exceed the conventional limit of $F = 4n^{2}$ across a broad range of wavelengths. We then numerically examine the effect of parasitic absorption across all absorption regimes by considering a plasmonic system with high local density of state: a metal-insulator-metal (MIM) waveguide. We show how the metal's material loss reduces the absorption enhancement in a thin core layer of a high-efficiency organic semiconductor, but still holds the potential to exceed conventional absorption enhancement limits.

\section{Statistical Coupled-Mode Theory for Multiple Lossy Channels}

Consider a resonance in a optical structure excited by an external plane wave. The incident plane wave comes from a particular channel, as specified, for example, by a particular angle of incidence. In addition, the resonance can couple to a total of $N$ different channels in free space, including and in addition to the channel where the incident wave is coming from. The resonance amplitude $a$ is then described by the temporal coupled mode formalism:
\begin{align}
\label{cmt}
\frac{d}{dt}a = \left ( i \omega_0 - \frac{N\gamma + \gamma_i + \gamma_p}{2} \right ) a + i \sqrt{\gamma}S
\end{align}
\ \\
Here $a$ is the resonance amplitude, $\omega_0$ the resonance frequency, $\gamma$ the external coupling rate, and $\gamma_i$ the intrinsic absorption rate in the \emph{active} material. We assume that the resonance has an isotropic response, i.e. its coupling rate to all external channels is the same at $\gamma$. Unlike previous formulations of light trapping and new to this work, we also introduce a parasitic absorption rate $\gamma_p$ which corresponds to modal loss in a non-active material. $S$ is the amplitude of the incident plane wave, with $|S|^2$ corresponding to its intensity.

We consider an incident wave at a particular frequency $\omega$, 
\begin{align}
S(t) &= S(\omega)\exp(i\omega t)
\end{align}
we then have
\begin{align}
a(t) &= a(\omega)\exp(i\omega t).
\end{align}
Substituting into Eq. \eqref{cmt} we find the following expression for the resonance amplitude $a$:
\begin{align}
a(\omega) = \frac{i\sqrt{\gamma} S(\omega)}{i (\omega - \omega_0 ) + (\gamma_i + \gamma_p + N\gamma)/2}.
\end{align}
This leads to separate expressions for absorption in the active layer $A_{act}$ and parasitic absorption $A_p$ due to the resonance:
\begin{align}
A_{act} &= \frac{\gamma_i \gamma}{(\omega - \omega_0 )^2 + (\gamma_i + \gamma_p + N\gamma)^2 /4}\\
A_p &= \frac{\gamma_p \gamma}{(\omega - \omega_0 )^2 + (\gamma_i + \gamma_p + N\gamma)^2 /4}
\end{align}

To characterize the contribution of the resonance to the broadband absorption enhancement, we then compute the corresponding spectral absorption cross section $\sigma_{act}$\cite{YuP} 
\begin{align}
\sigma_{act} = \int_{-\infty}^{\infty} A_{act}(\omega) d\omega = 2\pi\gamma_i \frac{1}{N + \gamma_i/\gamma + \gamma_p/\gamma}
\end{align}

A larger $\sigma_{act}$ corresponds to the stronger contribution of the resonance to the overall broad band absorption. To reach its maximum value of $\sigma_{max} = 2\pi\gamma_i / N$ for $\sigma_{act}$ the structure must operate in the \emph{overcoupling} regime where $\gamma \gg \gamma_i$ and $\gamma \gg \gamma_p$. The above analysis is for a single resonance in the absorbing structure. To get the total absorption coefficient $A$ one must sum over all $m$ resonances within a frequency range $\Delta \omega$:
\begin{align}
\label{firstA}
A = \frac{2\pi\gamma_i}{\Delta \omega} \sum_{m} \frac{\gamma_{m}}{N\gamma_{m} + \gamma_i + \gamma_p}
\end{align}
Ref. \cite{Yu2012} provides the following upper bound on the coupling rates $\gamma_{m}$, assuming $M$ resonances in the relevant bandwidth $[\omega, \omega + \Delta \omega]$ 
\begin{align}
\gamma_m \leq \frac{N}{M} \frac{\Delta\omega}{2\pi}.
\end{align}
Plugging this into Eq. \eqref{firstA} we find our first main result, which bounds the absorption in its most general way, in the presence of parasitic absorption:
\begin{align}
\label{first}
A \leq \frac{1}{1 + \gamma_p/\gamma_i + \frac{N}{M}\frac{\Delta\omega}{2\pi}}
\end{align}
To render this expression more readable we introduce two new terms, the single-pass absorption of the active material $\alpha_0 d$ and the raw enhancement factor $F_0$ for the structure in question in the weak-absorption limit and assuming no parasitic absorption. We can then rewrite Eq. \eqref{first} as
\begin{align}
\label{seconda}
A \leq \cfrac{\alpha_0 d}{\alpha_0 d \left (1 + \cfrac{\gamma_p}{\gamma_i} \right ) + \cfrac{1}{F_0}}
\end{align}
where
\begin{align}
F_0 = \frac{M}{N} \frac{2\pi\gamma_i}{\alpha_0 d \Delta \omega}.
\end{align}
In general, we emphasize that $F_0$ and $\gamma_p/\gamma_i$ are functions of wavelength in most realistic photovoltaics materials and light trapping schemes. This wavelength-dependence is important to consider, and thus the effect of parasitic absorption across all absorption regimes will be illustrated in the numerical example section with real material systems. 

Furthermore, previous work has indicated that if the parasitic material is a metal, $\gamma_p$ is subject an upper bound defined by the metal's material parameters\cite{Raman2013}. In general, we further observed that if the mode in the absorber has sub-wavelength volume, $\gamma_p$ is typically very near this upper bound. Specifically, if the metal is well described by the Drude model, then for strongly confined modes, such as surface modes near the surface plasmon frequency, $\gamma_p \sim \Gamma/2$ where $\Gamma$ is the Drude damping rate of the metal.

For an optical mode in an arbitrary nanostructure we can express $\gamma_i$ as $\gamma_i = \alpha^{wg}_{i} \cdot c / n_{wg}$, where $\alpha^{wg}_{i}$ is the the modal absorption coefficient in the active layer and $n_{wg}$ the group index of the mode. If we then define the modal overlap factor of the electric field intensity with the active material $V_{act} = \alpha^{wg}_i / \alpha_0$.

We now extend our analysis of Eq. \eqref{seconda} by first considering the case of weak active layer absorption which has been a focus of all light trapping designs, before considering regimes of arbitrary $\alpha_0 d$. An example of such a case is when plasmonic nanoparticles are used to enhancem absorption in the 800 - 1100 nm range for thin crystalline Silicon\cite{{catchpole2008},{schiff2011}}. To facilitate study of this scenario we define a modified enhancement factor $F_p$ in the weak-absorption regime (in the active layer) which does not assume $\gamma_p$ is negligible, as one does when deriving $F_0$. Beginning with Eq. \eqref{firstA} and assuming $\gamma \gg \gamma_i$ we find that
\begin{align}
\label{fp}
F_p &= \frac{M}{N} \frac{2\pi\gamma_i}{\alpha d \Delta \omega} \left ( 1 + \cfrac{M}{N}\cfrac{2\pi\gamma_p}{\Delta\omega} \right )^{-1}.
\end{align} 
This then allows us to re-express Eq. \eqref{seconda} in a manner analogous to the original upper bounds on absorption
\begin{align}
A \leq \cfrac{\alpha d}{\alpha d + \cfrac{1}{F_p}}.
\end{align}
One can then rewrite Eq. \eqref{fp} in terms of the original weak-absorption, non-parasitic enhancement factor $F_0$, $\alpha_0 d$ and $\gamma_p / \gamma_i$:
\begin{align}
\label{fp2}
F_p = \cfrac{F_0}{1+ \alpha_0 d F_0\cdot \cfrac{\gamma_p}{\gamma_i}} 
\end{align}
Eq. \eqref{fp2} shows that the raw enhancement factor is suppressed by the ratio of the modal absorption rate in the parasitic material, $\gamma_p$ to the active material's intrinsic absorption rate. Furthermore it reveals that the \emph{greater} the raw enhancement factor $F_0$ the more it will be suppressed by the presence of parasitic absorption. 

\section{Impact on Enhancement Factor}

To develop physical intuition into how the enhancement factor $F$ is reduced by the presence of parasitic absorption, we consider the simplest possible model: a parasitic absorber homogeneously distributed inside the active material.  In such a situation, where $n$ is the bulk refractive index of the composite material, $\gamma_p = \alpha_p \cdot c/n$ and $\gamma_i = \alpha \cdot c/n$. Eq. \eqref{fp2} then reduces to
\begin{align}
\label{simplefp2}
\frac{F_p}{F_0} = \frac{1}{1 + \alpha_p d F_0}.
\end{align}
We see immediately that the reduction of the raw enhancement factor $F_0$ is dependent on both the strength of the parasitic absorption $\alpha_p$ \emph{and} the raw enhancement factor $F_0$ itself. What this immediately suggests is that a light trapping design which provides a nominally high enhancement factor $F_0$ due to mechanisms such as field confinement or scattering is in fact \emph{more} sensitive to the presence of parasitic absorption than a design which provides a smaller enhancement factor. Thus, as we show in Fig. 1, design with higher raw enhancement factor but significant parasitic absorption can be worse than one with a significantly lower enhancement factor but lower parasitic absorption. 

The presence of the thickness $d$ is indicative of the fact that the amount the incident field interacts with the active and parasitic materials plays an important role in suppressing the enhancement factor, an issue we next consider in detail.

In most nanostructures of interest the parasitic component will not be evenly distributed in the manner previously considered. In such cases the real-space profile of the modes in question becomes essential to understand the effect of parasitic absorption. For an optical mode in an arbitrary nanostructure we can express $\gamma_i$ as $\gamma_i = \alpha^{wg}_{i} \cdot c / n_{wg}$, where $\alpha^{wg}_{i}$ is the the modal absorption coefficient in the active layer and $n_{wg}$ the group index of the mode. If we then define the modal overlap factor of the electric field intensity with the active material $V_{act} = \alpha^{wg}_i / \alpha_0$ we can re-write Eq. \eqref{simplefp2} as
\begin{align}
\label{simplefp3}
\frac{F_p}{F_0} = \cfrac{1}{1+ \gamma_p F_0 \left ( \cfrac{n_{wg}d}{cV_{act}}\right )}.
\end{align}
This expression is useful for complex photonic nanostructures since we can directly calculate $\gamma_p$, the imaginary frequency of the mode in the limit of $\gamma_p \gg \gamma_i$, using analytical and numerical techniques. 

Alternatively, for non-plasmonic parasitic absorbers, it is useful to express Eq. \eqref{simplefp3} in terms of the parasitic material's absorption coefficient. To do so we first define $\gamma_p = \alpha^{wg}_{p} \cdot c / n_{wg}$ where $\alpha^{wg}_{p}$ is the modal absorption coefficient of the parasitic material. We can now also define $V_{par} = \alpha^{wg}_p / \alpha_p$, the modal overlap factor of the electric field with the parasitic material. Eq. \eqref{simplefp3} then becomes 
\begin{align}
\label{simplefp4}
\frac{F_p}{F_0} = \cfrac{1}{1+ \alpha_p d F_0 \left ( \cfrac{V_{par}}{V_{act}}\right )}.
\end{align}
In the case of localized plasmon resonances, for example, where the field has equal intensity in the parasitic and active materials, $V_{par} = V_{act}$, and Eq. \eqref{simplefp4} reduces to Eq. \eqref{simplefp2}.

In Fig. 1 we examine how an increased modal overlap ratio reduces $F_p$ for different values of $F_0$ and $\alpha_p$. We observe that a larger absorption coefficient in the parasitic material, $\alpha_p$, has a greater effect on \emph{higher} enhancement-factor ($F_0$) achieving designs. Thus, a nanophotonic design that nominally achieves a greater $F_0$ but relies on a lossy parasitic material with strong modal overlap may perform only slightly better than a simpler design with smaller $F_0$ but low-loss non-active materials. We also note that, even in the weak absorption regime, parasitic loss does indeed suppress a nanophotonic structure's achievable absorption enhancement limit.

\begin{figure}[t]
 \begin{center}
   \includegraphics[scale=0.7]{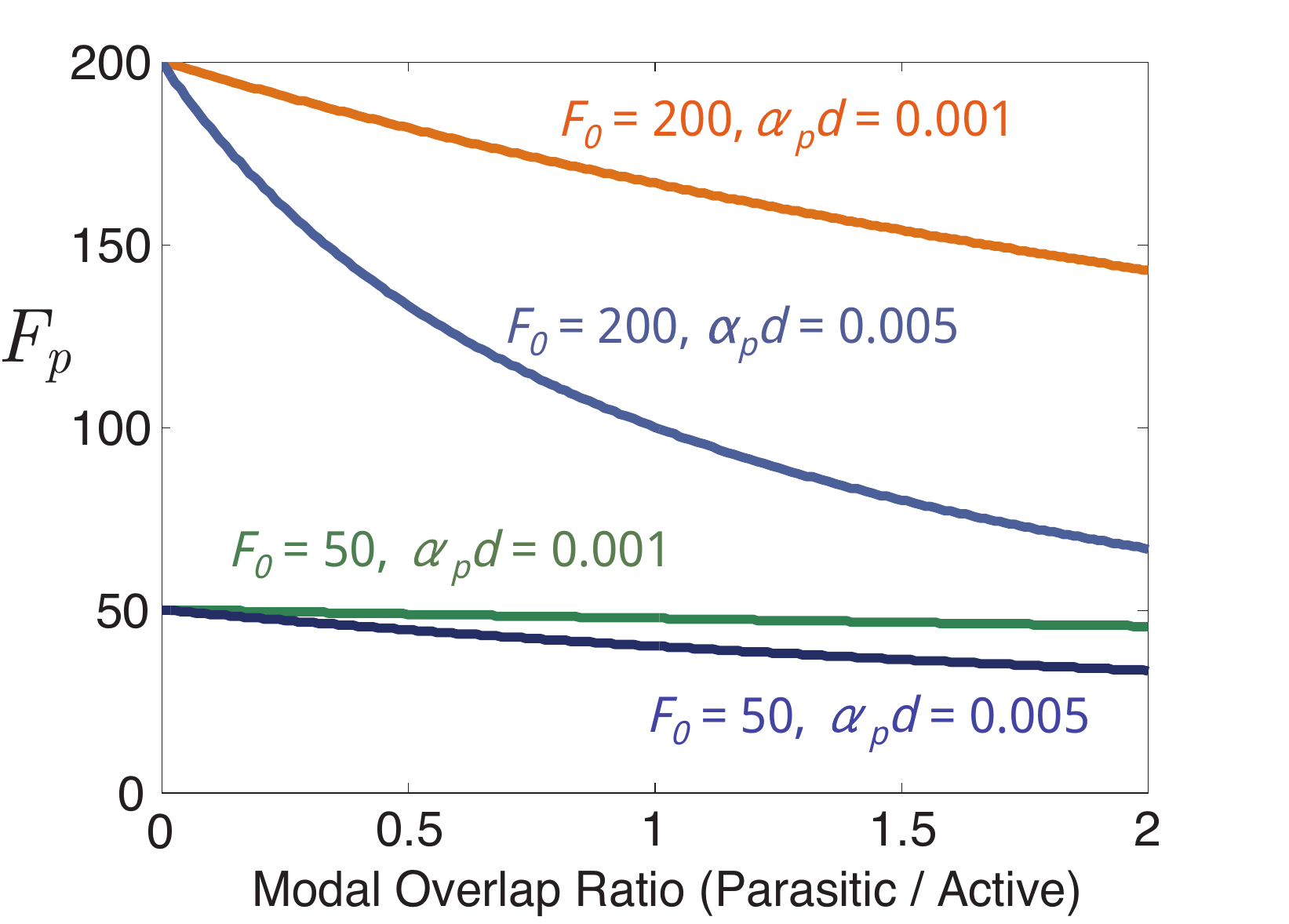}

  \end{center}
  \caption{The modified enhancement factor limit $F_p$ as a function of the modal overlap ratio between the parasitic and active material for systems of low and high idealized enhancement factors $F_0$, and small and large scaled parasitic absorption coefficients $\alpha_{p}d$.}
\end{figure}

The thin-film limit is a case of extreme interest for solar cells as it presents numerous opportunities to exceed the conventional $4n^2$ limit\cite{{YuP},{munday2012}} and use less material to generate power equivalent to thicker cells. Moreover current state-of-the-art nanophotonic approaches are targeted and hold the greatest promise for thin, sub-wavelength active layer thicknesses. Previously, it was shown that the raw enhancement factor $F_0$ for each mode in the thin-film case is $F^{mode}_0 = \frac{\lambda}{d} n_{wg} V_{act}$\cite{YuP}. We can substitute this into Eq. \eqref{simplefp3} to find a modified modal enhancement factor 
\begin{equation}
\label{nanofp}
F^{mode}_{p} = \cfrac{\cfrac{\lambda}{d} n_{wg} V_{act}}{1 + \gamma_{p} V_{act} n^{2}_{wg} \cfrac{\lambda}{c}}.
\end{equation}
As discussed earlier, if the parasitic material is assumed to be a Drude metal with damping rate $\Gamma$, and when there is strong field confinement, one can substitute $\gamma_p = \Gamma/2$ in Eq. \eqref{nanofp}. Similarly one can substitute into Eq. \eqref{simplefp4} to find a modified modal enhancement factor written in terms of the parasitic material's absorption coefficient
\begin{equation}
F^{mode}_{p} = \cfrac{\cfrac{\lambda}{d} n_{wg} V_{act}}{1 + \alpha_{p} \lambda n_{wg} V_{par}}. 
\end{equation}

To determine the overall value for $F_p$, one would then count the equivalent adjusted enhancement from each mode present, i.e. $F_p = \sum_{modes} F_p^{mode}$. The important feature of these equations is that it rigorously calculates the enhancement factor limit for potential beyond-$4n^2$ systems, even in the presence of parasitic loss. With it, we also now have a rigorous connection between $F_p$ and the absorption coefficient of the parasitic material $\alpha_p$ and the modal overlap factor $V_{par}$. 

\section{Numerical Study of a Canonical Plasmonic System}

\begin{figure}[t]
 \begin{center}
   \includegraphics[scale=0.7]{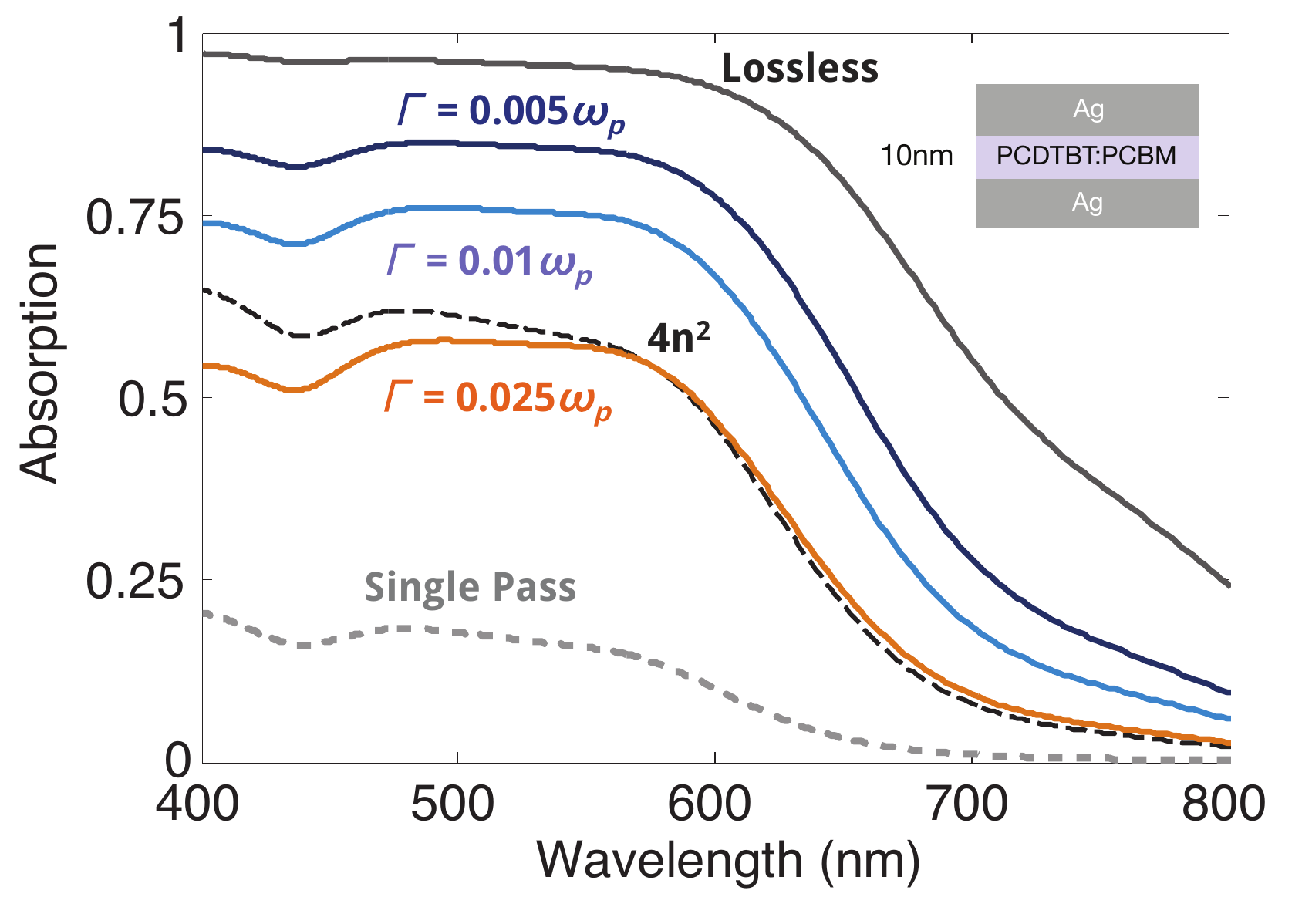}

  \end{center}
  \caption{The modal absorption limit for a MIM waveguide with a 10 nm active layer of PCDTBT:PC$_{70}BM$ and where silver's Drude plasma frequency defines the metal. As the metal's Drude damping rate is increased the absorption limit across all wavelengths is reduced, but still performs well relative to the conventional $4n^{2}$ limit.}
\end{figure}

To capture the effect of parasitic loss on a light trapping scheme across all values of the active layer's absorption coefficient, we consider a canonical plasmonic system: a metal-insulator-metal (MIM) waveguide where the Drude metal has silver's plasma frequency and a variable damping rate $\Gamma$, and the dielectric core/ active layer is a high-efficiency organic bulk-heterojunction semiconductor PCDTBT:PC$_{70}$BM $(n\sim 2)$\cite{{Park2009},{Raman2011}}. Since such organic materials need to be deposited in thin layers for efficient carrier extraction, we consider a very thin active layer thickness of 10 nm. The entire structure then supports the fundamental gap-plasmon mode across the entire relevant wavelength range, where we consider the resulting absorption enhancement.

To determine the absorption limit we first use Eq. \eqref{nanofp} to find $F_p (\lambda)$. To do so, we directly calculate\cite{kekatpure2009} the parasitic modal loss rate $\gamma_p$, modal overlap factor $V_{act}$ and mode index $n_{wg}$ of the fundamental gap plasmon mode at all relevant wavelengths with actual material parameters at those wavelengths. The calculated $F_p (\lambda)$, $\gamma_p (\lambda)$ and $\gamma_i (\lambda)$ are then substituted into Eq. \eqref{seconda} to calculate the modal absorption limit. In Fig. 2 we plot the absorption limit $A(\lambda)$ for this mode as calculated for varying values of the Drude damping rate in the metal.  We emphasize that this limit is a modal calculation, and assumes ideal coupling into the fundamental gap plasmon mode, which in practice can be challenging in the geometry shown. We observe that realistic values of $\Gamma \sim 0.01\omega_p$ suppress the absorption limit below the idealized lossless case, but it remains above the conventional $4n^{2}$ limit. Stronger $\Gamma$ values however can cause the enhancement to go below the conventional limit, indicating the need for careful material choice and nanostructure design in using plasmonics for light trapping.

Finally, to illustrate the how parasitic absorption affects limits for varying thicknesses of the active layer, we examine the modal absorption limit for the MIM waveguide scenario in Fig. 3. We fix the parasitic loss rate $\gamma_p$ at its worst value, $\Gamma/2$\cite{Raman2013}, and consider two limits of the active layer thickness, 5 nm and 100 nm. Even with maximal parasitic loss in the metal, the modal confinement for the 5 nm thick active layer is sufficient to far exceed the conventional $4n^2$ limit. However, with a thicker active layer the parasitic absorption actually brings the modal absorption limit \emph{below} the conventional limit. This indicates that light trapping beyond the conventional limit is possible for very thin active layers even in the presence of very high parasitic losses. But, as one goes to thicker active layers, the effect of parasitic losses can outweigh the added benefit from modal confinement in terms of enhancement factors.

\begin{figure}[t]
 \begin{center}
   \includegraphics[scale=0.7]{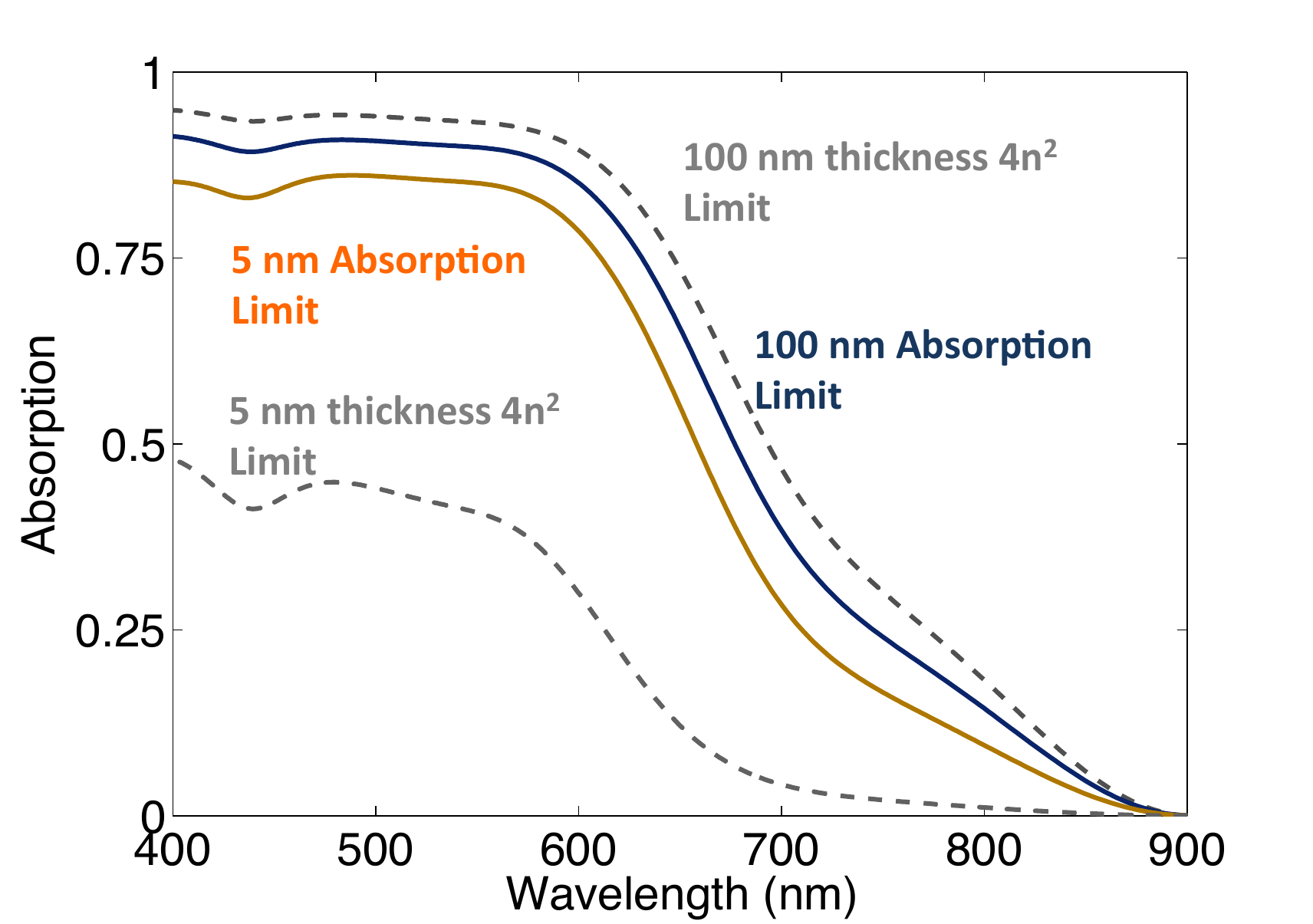}

  \end{center}
  \caption{The modal absorption limit for a MIM waveguide with the worst-case parasitic loss rate of $\gamma_p = \Gamma/2$ for two scenarios: a 5 nm active layer and a 100 nm active layer of PCDTBT:PC$_{70}BM$. A Drude fit of silver's dielectric function defines the metal. For the 5 nm thick case, the absorption limit for the fundamental MIM mode with worst-case parasitic loss is substantially higher than the corresponding $4n^2$ limit, whereas for the 100 nm case, the limit with worst-case loss is in fact below the $4n^2$ limit. }
\end{figure}

We have thus rigorously derived a theory on the effect of multiple lossy materials in any broadband photonic absorption enhancement scheme across all absorption regimes. These results indicate that, while parasitic loss in a non-active material can suppress the achievable absorption enhancement, conventional plasmonic schemes can still exceed conventional limits on light trapping with appropriate design and material selection. While we have focused on broadband light trapping for solar applications as a motivating scenario, these results point to a wide range of opportunities that may lie in studying the interaction of electromagnetic modes with multiple lossy materials in the same nanophotonic structure.

\begin{acknowledgement}
S. F. acknowledges the support of Department of Energy Grant No. DE-FG02-07ER46426
\end{acknowledgement}

\providecommand{\latin}[1]{#1}
\makeatletter
\providecommand{\doi}
  {\begingroup\let\do\@makeother\dospecials
  \catcode`\{=1 \catcode`\}=2\doi@aux}
\providecommand{\doi@aux}[1]{\endgroup\texttt{#1}}
\makeatother
\providecommand*\mcitethebibliography{\thebibliography}
\csname @ifundefined\endcsname{endmcitethebibliography}
  {\let\endmcitethebibliography\endthebibliography}{}


\begin{mcitethebibliography}{42}
\providecommand*\natexlab[1]{#1}
\providecommand*\mciteSetBstSublistMode[1]{}
\providecommand*\mciteSetBstMaxWidthForm[2]{}
\providecommand*\mciteBstWouldAddEndPuncttrue
  {\def\EndOfBibitem{\unskip.}}
\providecommand*\mciteBstWouldAddEndPunctfalse
  {\let\EndOfBibitem\relax}
\providecommand*\mciteSetBstMidEndSepPunct[3]{}
\providecommand*\mciteSetBstSublistLabelBeginEnd[3]{}
\providecommand*\EndOfBibitem{}
\mciteSetBstSublistMode{f}
\mciteSetBstMaxWidthForm{subitem}{(\alph{mcitesubitemcount})}
\mciteSetBstSublistLabelBeginEnd
  {\mcitemaxwidthsubitemform\space}
  {\relax}
  {\relax}
\bibitem[Yablonovitch(1982)]{Yab82}
Yablonovitch,~E. Statistical ray optics. \emph{Journal of the Optical Society
  of America} \textbf{1982}, \emph{72}, 899--907\relax
\mciteBstWouldAddEndPuncttrue
\mciteSetBstMidEndSepPunct{\mcitedefaultmidpunct}
{\mcitedefaultendpunct}{\mcitedefaultseppunct}\relax
\EndOfBibitem
\bibitem[Yablonovitch and Cody(1982)Yablonovitch, and Cody]{yablonovitch1982}
Yablonovitch,~E.; Cody,~G.~D. Intensity enhancement in textured optical sheets
  for solar cells. \emph{IEEE Transactions on Electron Devices} \textbf{1982},
  \emph{29}, 300--305\relax
\mciteBstWouldAddEndPuncttrue
\mciteSetBstMidEndSepPunct{\mcitedefaultmidpunct}
{\mcitedefaultendpunct}{\mcitedefaultseppunct}\relax
\EndOfBibitem
\bibitem[Stuart and Hall(1997)Stuart, and Hall]{StuartHall}
Stuart,~H.~R.; Hall,~D.~G. Thermodynamic limit to light trapping in thin planar
  structures. \emph{Journal of the Optical Society of America A} \textbf{1997},
  \emph{14}, 3001--3008\relax
\mciteBstWouldAddEndPuncttrue
\mciteSetBstMidEndSepPunct{\mcitedefaultmidpunct}
{\mcitedefaultendpunct}{\mcitedefaultseppunct}\relax
\EndOfBibitem
\bibitem[Zhu \latin{et~al.}(2009)Zhu, Hsu, Yu, Fan, and Cui]{zhu2009}
Zhu,~J.; Hsu,~C.-M.; Yu,~Z.; Fan,~S.; Cui,~Y. Nanodome solar cells with
  efficient light management and self-cleaning. \emph{Nano Letters}
  \textbf{2009}, \emph{10}, 1979--1984\relax
\mciteBstWouldAddEndPuncttrue
\mciteSetBstMidEndSepPunct{\mcitedefaultmidpunct}
{\mcitedefaultendpunct}{\mcitedefaultseppunct}\relax
\EndOfBibitem
\bibitem[Mallick \latin{et~al.}(2010)Mallick, Agrawal, and Peumans]{Mallick10}
Mallick,~S.~B.; Agrawal,~M.; Peumans,~P. Optimal light trapping in ultra-thin
  photonic crystal crystalline silicon solar cells. \emph{Optics Express}
  \textbf{2010}, \emph{18}, 5691--5706\relax
\mciteBstWouldAddEndPuncttrue
\mciteSetBstMidEndSepPunct{\mcitedefaultmidpunct}
{\mcitedefaultendpunct}{\mcitedefaultseppunct}\relax
\EndOfBibitem
\bibitem[Garnett and Yang(2010)Garnett, and Yang]{garnett2010}
Garnett,~E.; Yang,~P. Light trapping in silicon nanowire solar cells.
  \emph{Nano Letters} \textbf{2010}, \emph{10}, 1082--1087\relax
\mciteBstWouldAddEndPuncttrue
\mciteSetBstMidEndSepPunct{\mcitedefaultmidpunct}
{\mcitedefaultendpunct}{\mcitedefaultseppunct}\relax
\EndOfBibitem
\bibitem[Ferry \latin{et~al.}(2010)Ferry, Verschuuren, Li, Verhagen, Walters,
  Schropp, Atwater, and Polman]{ferry2010}
Ferry,~V.~E.; Verschuuren,~M.~A.; Li,~H.~B.; Verhagen,~E.; Walters,~R.~J.;
  Schropp,~R.~E.; Atwater,~H.~A.; Polman,~A. Light trapping in ultrathin
  plasmonic solar cells. \emph{Optics Express} \textbf{2010}, \emph{18},
  A237--A245\relax
\mciteBstWouldAddEndPuncttrue
\mciteSetBstMidEndSepPunct{\mcitedefaultmidpunct}
{\mcitedefaultendpunct}{\mcitedefaultseppunct}\relax
\EndOfBibitem
\bibitem[Polman and Atwater(2012)Polman, and Atwater]{polman2012}
Polman,~A.; Atwater,~H.~A. Photonic design principles for ultrahigh-efficiency
  photovoltaics. \emph{Nature Materials} \textbf{2012}, \emph{11},
  174--177\relax
\mciteBstWouldAddEndPuncttrue
\mciteSetBstMidEndSepPunct{\mcitedefaultmidpunct}
{\mcitedefaultendpunct}{\mcitedefaultseppunct}\relax
\EndOfBibitem
\bibitem[Battaglia \latin{et~al.}(2012)Battaglia, Hsu, Soderstrom, Escarre,
  Haug, Charriere, Boccard, Despeisse, Alexander, Cantoni, \latin{et~al.}
  others]{battaglia2012}
others,, \latin{et~al.}  Light trapping in solar cells: can periodic beat
  random? \emph{ACS Nano} \textbf{2012}, \emph{6}, 2790--2797\relax
\mciteBstWouldAddEndPuncttrue
\mciteSetBstMidEndSepPunct{\mcitedefaultmidpunct}
{\mcitedefaultendpunct}{\mcitedefaultseppunct}\relax
\EndOfBibitem
\bibitem[Grote \latin{et~al.}(2013)Grote, Brown, Driscoll, Osgood, and
  Schuller]{grote2013}
Grote,~R.~R.; Brown,~S.~J.; Driscoll,~J.~B.; Osgood,~R.~M.; Schuller,~J.~A.
  Morphology-dependent light trapping in thin-film organic solar cells.
  \emph{Optics Express} \textbf{2013}, \emph{21}, A847--A863\relax
\mciteBstWouldAddEndPuncttrue
\mciteSetBstMidEndSepPunct{\mcitedefaultmidpunct}
{\mcitedefaultendpunct}{\mcitedefaultseppunct}\relax
\EndOfBibitem
\bibitem[Narasimhan and Cui(2013)Narasimhan, and Cui]{narasimhan2013}
Narasimhan,~V.~K.; Cui,~Y. Nanostructures for photon management in solar cells.
  \emph{Nanophotonics} \textbf{2013}, \emph{2}, 187--210\relax
\mciteBstWouldAddEndPuncttrue
\mciteSetBstMidEndSepPunct{\mcitedefaultmidpunct}
{\mcitedefaultendpunct}{\mcitedefaultseppunct}\relax
\EndOfBibitem
\bibitem[Brongersma \latin{et~al.}(2014)Brongersma, Cui, and
  Fan]{brongersma2014}
Brongersma,~M.~L.; Cui,~Y.; Fan,~S. Light management for photovoltaics using
  high-index nanostructures. \emph{Nature Materials} \textbf{2014}, \emph{13},
  451--460\relax
\mciteBstWouldAddEndPuncttrue
\mciteSetBstMidEndSepPunct{\mcitedefaultmidpunct}
{\mcitedefaultendpunct}{\mcitedefaultseppunct}\relax
\EndOfBibitem
\bibitem[Yu \latin{et~al.}(2010)Yu, Raman, and Fan]{YuP}
Yu,~Z.; Raman,~A.; Fan,~S. {Fundamental limit of nanophotonic light trapping in
  solar cells}. \emph{Proceedings of the National Academy of Sciences}
  \textbf{2010}, \emph{107}, 17491--17496\relax
\mciteBstWouldAddEndPuncttrue
\mciteSetBstMidEndSepPunct{\mcitedefaultmidpunct}
{\mcitedefaultendpunct}{\mcitedefaultseppunct}\relax
\EndOfBibitem
\bibitem[Yu \latin{et~al.}(2010)Yu, Raman, and Fan]{YuOE}
Yu,~Z.; Raman,~A.; Fan,~S. Fundamental limit of light trapping in grating
  structures. \emph{Optics Express} \textbf{2010}, \emph{18}, A366--A380\relax
\mciteBstWouldAddEndPuncttrue
\mciteSetBstMidEndSepPunct{\mcitedefaultmidpunct}
{\mcitedefaultendpunct}{\mcitedefaultseppunct}\relax
\EndOfBibitem
\bibitem[Callahan \latin{et~al.}(2012)Callahan, Munday, and
  Atwater]{Callahan2012}
Callahan,~D.~M.; Munday,~J.~N.; Atwater,~H.~A. Solar Cell Light Trapping beyond
  the Ray Optic Limit. \emph{Nano Letters} \textbf{2012}, \emph{12},
  214--218\relax
\mciteBstWouldAddEndPuncttrue
\mciteSetBstMidEndSepPunct{\mcitedefaultmidpunct}
{\mcitedefaultendpunct}{\mcitedefaultseppunct}\relax
\EndOfBibitem
\bibitem[Yu \latin{et~al.}(2012)Yu, Raman, and Fan]{Yu2012}
Yu,~Z.; Raman,~A.; Fan,~S. Thermodynamic upper bound on broadband light
  coupling with photonic structures. \emph{Physical Review Letters}
  \textbf{2012}, \emph{109}, 173901\relax
\mciteBstWouldAddEndPuncttrue
\mciteSetBstMidEndSepPunct{\mcitedefaultmidpunct}
{\mcitedefaultendpunct}{\mcitedefaultseppunct}\relax
\EndOfBibitem
\bibitem[Schiff(2011)]{schiff2011}
Schiff,~E.~A. Thermodynamic limit to photonic-plasmonic light-trapping in thin
  films on metals. \emph{Journal of Applied Physics} \textbf{2011}, \emph{110},
  104501\relax
\mciteBstWouldAddEndPuncttrue
\mciteSetBstMidEndSepPunct{\mcitedefaultmidpunct}
{\mcitedefaultendpunct}{\mcitedefaultseppunct}\relax
\EndOfBibitem
\bibitem[Munday \latin{et~al.}(2012)Munday, Callahan, and Atwater]{munday2012}
Munday,~J.~N.; Callahan,~D.~M.; Atwater,~H.~A. Light trapping beyond the 4n[sup
  2] limit in thin waveguides. \emph{Applied Physics Letters} \textbf{2012},
  \emph{100}, 121121\relax
\mciteBstWouldAddEndPuncttrue
\mciteSetBstMidEndSepPunct{\mcitedefaultmidpunct}
{\mcitedefaultendpunct}{\mcitedefaultseppunct}\relax
\EndOfBibitem
\bibitem[Mokkapati and Catchpole(2012)Mokkapati, and Catchpole]{mokkapati2012}
Mokkapati,~S.; Catchpole,~K. Nanophotonic light trapping in solar cells.
  \emph{Journal of Applied Physics} \textbf{2012}, \emph{112}, 101101\relax
\mciteBstWouldAddEndPuncttrue
\mciteSetBstMidEndSepPunct{\mcitedefaultmidpunct}
{\mcitedefaultendpunct}{\mcitedefaultseppunct}\relax
\EndOfBibitem
\bibitem[Buddhiraju and Fan(2017)Buddhiraju, and Fan]{buddhiraju2017}
Buddhiraju,~S.; Fan,~S. Theory of solar cell light trapping through a
  nonequilibrium Green's function formulation of Maxwell's equations.
  \emph{Physical Review B} \textbf{2017}, \emph{96}, 035304\relax
\mciteBstWouldAddEndPuncttrue
\mciteSetBstMidEndSepPunct{\mcitedefaultmidpunct}
{\mcitedefaultendpunct}{\mcitedefaultseppunct}\relax
\EndOfBibitem
\bibitem[Catchpole and Polman(2008)Catchpole, and Polman]{catchpole2008}
Catchpole,~K.; Polman,~A. Plasmonic solar cells. \emph{Optics Express}
  \textbf{2008}, \emph{16}, 21793--21800\relax
\mciteBstWouldAddEndPuncttrue
\mciteSetBstMidEndSepPunct{\mcitedefaultmidpunct}
{\mcitedefaultendpunct}{\mcitedefaultseppunct}\relax
\EndOfBibitem
\bibitem[Kim \latin{et~al.}(2008)Kim, Na, Jo, Kim, and Nah]{kim2008plasmon}
Kim,~S.-S.; Na,~S.-I.; Jo,~J.; Kim,~D.-Y.; Nah,~Y.-C. Plasmon enhanced
  performance of organic solar cells using electrodeposited Ag nanoparticles.
  \emph{Applied Physics Letters} \textbf{2008}, \emph{93}, 073307\relax
\mciteBstWouldAddEndPuncttrue
\mciteSetBstMidEndSepPunct{\mcitedefaultmidpunct}
{\mcitedefaultendpunct}{\mcitedefaultseppunct}\relax
\EndOfBibitem
\bibitem[Beck \latin{et~al.}(2009)Beck, Polman, and Catchpole]{beck2009}
Beck,~F.; Polman,~A.; Catchpole,~K. Tunable light trapping for solar cells
  using localized surface plasmons. \emph{Journal of Applied Physics}
  \textbf{2009}, \emph{105}, 114310\relax
\mciteBstWouldAddEndPuncttrue
\mciteSetBstMidEndSepPunct{\mcitedefaultmidpunct}
{\mcitedefaultendpunct}{\mcitedefaultseppunct}\relax
\EndOfBibitem
\bibitem[Atwater and Polman(2010)Atwater, and Polman]{atwater2010}
Atwater,~H.~A.; Polman,~A. Plasmonics for improved photovoltaic devices.
  \emph{Nature Materials} \textbf{2010}, \emph{9}, 205--213\relax
\mciteBstWouldAddEndPuncttrue
\mciteSetBstMidEndSepPunct{\mcitedefaultmidpunct}
{\mcitedefaultendpunct}{\mcitedefaultseppunct}\relax
\EndOfBibitem
\bibitem[Ferry \latin{et~al.}(2010)Ferry, Munday, and Atwater]{ferry2010design}
Ferry,~V.~E.; Munday,~J.~N.; Atwater,~H.~A. Design considerations for plasmonic
  photovoltaics. \emph{Advanced Materials} \textbf{2010}, \emph{22},
  4794--4808\relax
\mciteBstWouldAddEndPuncttrue
\mciteSetBstMidEndSepPunct{\mcitedefaultmidpunct}
{\mcitedefaultendpunct}{\mcitedefaultseppunct}\relax
\EndOfBibitem
\bibitem[Green and Pillai(2012)Green, and Pillai]{Green2012}
Green,~M.~A.; Pillai,~S. Harnessing plasmonics for solar cells. \emph{Nature
  Photonics} \textbf{2012}, \emph{6}, 130--132\relax
\mciteBstWouldAddEndPuncttrue
\mciteSetBstMidEndSepPunct{\mcitedefaultmidpunct}
{\mcitedefaultendpunct}{\mcitedefaultseppunct}\relax
\EndOfBibitem
\bibitem[Spinelli \latin{et~al.}(2012)Spinelli, Ferry, Van~de Groep, Van~Lare,
  Verschuuren, Schropp, Atwater, and Polman]{spinelli2012}
Spinelli,~P.; Ferry,~V.; Van~de Groep,~J.; Van~Lare,~M.; Verschuuren,~M.;
  Schropp,~R.; Atwater,~H.; Polman,~A. Plasmonic light trapping in thin-film Si
  solar cells. \emph{Journal of Optics} \textbf{2012}, \emph{14}, 024002\relax
\mciteBstWouldAddEndPuncttrue
\mciteSetBstMidEndSepPunct{\mcitedefaultmidpunct}
{\mcitedefaultendpunct}{\mcitedefaultseppunct}\relax
\EndOfBibitem
\bibitem[Pala \latin{et~al.}(2009)Pala, White, Barnard, Liu, and
  Brongersma]{pala2009}
Pala,~R.~A.; White,~J.; Barnard,~E.; Liu,~J.; Brongersma,~M.~L. Design of
  plasmonic thin-film solar cells with broadband absorption enhancements.
  \emph{Advanced Materials} \textbf{2009}, \emph{21}, 3504--3509\relax
\mciteBstWouldAddEndPuncttrue
\mciteSetBstMidEndSepPunct{\mcitedefaultmidpunct}
{\mcitedefaultendpunct}{\mcitedefaultseppunct}\relax
\EndOfBibitem
\bibitem[Pillai and Green(2010)Pillai, and Green]{pillai2010}
Pillai,~S.; Green,~M. Plasmonics for photovoltaic applications. \emph{Solar
  Energy Materials and Solar Cells} \textbf{2010}, \emph{94}, 1481--1486\relax
\mciteBstWouldAddEndPuncttrue
\mciteSetBstMidEndSepPunct{\mcitedefaultmidpunct}
{\mcitedefaultendpunct}{\mcitedefaultseppunct}\relax
\EndOfBibitem
\bibitem[Ferry \latin{et~al.}(2011)Ferry, Polman, and
  Atwater]{ferry2011modeling}
Ferry,~V.~E.; Polman,~A.; Atwater,~H.~A. Modeling light trapping in
  nanostructured solar cells. \emph{ACS nano} \textbf{2011}, \emph{5},
  10055--10064\relax
\mciteBstWouldAddEndPuncttrue
\mciteSetBstMidEndSepPunct{\mcitedefaultmidpunct}
{\mcitedefaultendpunct}{\mcitedefaultseppunct}\relax
\EndOfBibitem
\bibitem[Tan \latin{et~al.}(2012)Tan, Santbergen, Smets, and Zeman]{tan2012}
Tan,~H.; Santbergen,~R.; Smets,~A.~H.; Zeman,~M. Plasmonic light trapping in
  thin-film silicon solar cells with improved self-assembled silver
  nanoparticles. \emph{Nano letters} \textbf{2012}, \emph{12}, 4070--4076\relax
\mciteBstWouldAddEndPuncttrue
\mciteSetBstMidEndSepPunct{\mcitedefaultmidpunct}
{\mcitedefaultendpunct}{\mcitedefaultseppunct}\relax
\EndOfBibitem
\bibitem[Pala \latin{et~al.}(2013)Pala, Liu, Barnard, Askarov, Garnett, Fan,
  and Brongersma]{pala2013}
Pala,~R.~A.; Liu,~J.~S.; Barnard,~E.~S.; Askarov,~D.; Garnett,~E.~C.; Fan,~S.;
  Brongersma,~M.~L. Optimization of non-periodic plasmonic light-trapping
  layers for thin-film solar cells. \emph{Nature communications} \textbf{2013},
  \emph{4}, ncomms3095\relax
\mciteBstWouldAddEndPuncttrue
\mciteSetBstMidEndSepPunct{\mcitedefaultmidpunct}
{\mcitedefaultendpunct}{\mcitedefaultseppunct}\relax
\EndOfBibitem
\bibitem[Holman \latin{et~al.}(2014)Holman, Filipi{\v{c}}, Lipov{\v{s}}ek,
  De~Wolf, Smole, Topi{\v{c}}, and Ballif]{holman2014}
Holman,~Z.~C.; Filipi{\v{c}},~M.; Lipov{\v{s}}ek,~B.; De~Wolf,~S.; Smole,~F.;
  Topi{\v{c}},~M.; Ballif,~C. Parasitic absorption in the rear reflector of a
  silicon solar cell: Simulation and measurement of the sub-bandgap reflectance
  for common dielectric/metal reflectors. \emph{Solar Energy Materials and
  Solar Cells} \textbf{2014}, \emph{120}, 426--430\relax
\mciteBstWouldAddEndPuncttrue
\mciteSetBstMidEndSepPunct{\mcitedefaultmidpunct}
{\mcitedefaultendpunct}{\mcitedefaultseppunct}\relax
\EndOfBibitem
\bibitem[Morawiec \latin{et~al.}(2016)Morawiec, Holovsk{\`y}, Mendes,
  M{\"u}ller, Ganzerov{\'a}, Vetushka, Ledinsk{\`y}, Priolo, Fejfar, and
  Crupi]{morawiec2016}
Morawiec,~S.; Holovsk{\`y},~J.; Mendes,~M.~J.; M{\"u}ller,~M.;
  Ganzerov{\'a},~K.; Vetushka,~A.; Ledinsk{\`y},~M.; Priolo,~F.; Fejfar,~A.;
  Crupi,~I. Experimental quantification of useful and parasitic absorption of
  light in plasmon-enhanced thin silicon films for solar cells application.
  \emph{Scientific reports} \textbf{2016}, \emph{6}, 22481\relax
\mciteBstWouldAddEndPuncttrue
\mciteSetBstMidEndSepPunct{\mcitedefaultmidpunct}
{\mcitedefaultendpunct}{\mcitedefaultseppunct}\relax
\EndOfBibitem
\bibitem[Disney \latin{et~al.}(2017)Disney, Pillai, and Green]{disney2017}
Disney,~C.~E.; Pillai,~S.; Green,~M.~A. The Impact of parasitic loss on solar
  cells with plasmonic nano-textured rear reflectors. \emph{Scientific reports}
  \textbf{2017}, \emph{7}, 12826\relax
\mciteBstWouldAddEndPuncttrue
\mciteSetBstMidEndSepPunct{\mcitedefaultmidpunct}
{\mcitedefaultendpunct}{\mcitedefaultseppunct}\relax
\EndOfBibitem
\bibitem[Sheng \latin{et~al.}(2012)Sheng, Hu, Michel, and Kimerling]{sheng2012}
Sheng,~X.; Hu,~J.; Michel,~J.; Kimerling,~L.~C. Light trapping limits in
  plasmonic solar cells: an analytical investigation. \emph{Optics Express}
  \textbf{2012}, \emph{20}, A496--A501\relax
\mciteBstWouldAddEndPuncttrue
\mciteSetBstMidEndSepPunct{\mcitedefaultmidpunct}
{\mcitedefaultendpunct}{\mcitedefaultseppunct}\relax
\EndOfBibitem
\bibitem[Raman \latin{et~al.}(2014)Raman, Anoma, Zhu, Rephaeli, and
  Fan]{Raman2014}
Raman,~A.~P.; Anoma,~M.~A.; Zhu,~L.; Rephaeli,~E.; Fan,~S. {Passive radiative
  cooling below ambient air temperature under direct sunlight}. \emph{Nature}
  \textbf{2014}, \emph{515}, 540--544\relax
\mciteBstWouldAddEndPuncttrue
\mciteSetBstMidEndSepPunct{\mcitedefaultmidpunct}
{\mcitedefaultendpunct}{\mcitedefaultseppunct}\relax
\EndOfBibitem
\bibitem[Raman \latin{et~al.}(2013)Raman, Shin, and Fan]{Raman2013}
Raman,~A.; Shin,~W.; Fan,~S. Upper Bound on the Modal Material Loss Rate in
  Plasmonic and Metamaterial Systems. \emph{Physical Review Letters}
  \textbf{2013}, \emph{110}, 183901\relax
\mciteBstWouldAddEndPuncttrue
\mciteSetBstMidEndSepPunct{\mcitedefaultmidpunct}
{\mcitedefaultendpunct}{\mcitedefaultseppunct}\relax
\EndOfBibitem
\bibitem[Park \latin{et~al.}(2009)Park, Roy, Beaupre, Cho, Coates, Moon, Moses,
  Leclerc, Lee, and Heeger]{Park2009}
Park,~S.~H.; Roy,~A.; Beaupre,~S.; Cho,~S.; Coates,~N.; Moon,~J.~S.; Moses,~D.;
  Leclerc,~M.; Lee,~K.; Heeger,~A.~J. Bulk heterojunction solar cells with
  internal quantum efficiency approaching 100\%. \emph{Nature Photonics}
  \textbf{2009}, \emph{3}, 297 -- 302\relax
\mciteBstWouldAddEndPuncttrue
\mciteSetBstMidEndSepPunct{\mcitedefaultmidpunct}
{\mcitedefaultendpunct}{\mcitedefaultseppunct}\relax
\EndOfBibitem
\bibitem[Raman \latin{et~al.}(2011)Raman, Yu, and Fan]{Raman2011}
Raman,~A.; Yu,~Z.; Fan,~S. Dielectric nanostructures for broadband light
  trapping in organic solar cells. \emph{Optics Express} \textbf{2011},
  \emph{19}, 19015--19026\relax
\mciteBstWouldAddEndPuncttrue
\mciteSetBstMidEndSepPunct{\mcitedefaultmidpunct}
{\mcitedefaultendpunct}{\mcitedefaultseppunct}\relax
\EndOfBibitem
\bibitem[Kekatpure \latin{et~al.}(2009)Kekatpure, Hryciw, Barnard, and
  Brongersma]{kekatpure2009}
Kekatpure,~R.~D.; Hryciw,~A.~C.; Barnard,~E.~S.; Brongersma,~M.~L. Solving
  dielectric and plasmonic waveguide dispersion relations on a pocket
  calculator. \emph{Optics Express} \textbf{2009}, \emph{17},
  24112--24129\relax
\mciteBstWouldAddEndPuncttrue
\mciteSetBstMidEndSepPunct{\mcitedefaultmidpunct}
{\mcitedefaultendpunct}{\mcitedefaultseppunct}\relax
\EndOfBibitem
\end{mcitethebibliography}
\end{document}